\documentclass[preprint]{revtex4}
\usepackage{color}
\usepackage{graphicx}

\newcommand{\diff}{{\mathrm d}}

\begin{document}

\title{Ripening Kinetics of Bubbles: A Molecular Dynamics Study}

\author{Hiroshi Watanabe}
\email{hwatanabe@issp.u-tokyo.ac.jp}
\thanks{Corresponding author}
\affiliation{
The Institute for Solid State Physics, The University of Tokyo,
Kashiwanoha 5-1-5, Kashiwa, Chiba 277-8581, Japan
}

\author{Hajime Inaoka}
\affiliation{
Advanced Institute for Computational Science, RIKEN,
7-1-26, Minatojima-minami-machi, Chuo-ku, Kobe Hyogo, 650-0047, Japan
}

\author{Nobuyasu Ito}
\affiliation{
Department of Applied Physics, School of Engineering,
The University of Tokyo, Hongo, Bunkyo-ku, Tokyo 113-8656, Japan
}
\affiliation{
Advanced Institute for Computational Science, RIKEN,
7-1-26, Minatojima-minami-machi, Chuo-ku, Kobe, Hyogo 650-0047, Japan
}

\date{\today}

%\pacs{47.55.dp, 02.70.Ns, 65.20.De}

\begin{abstract}
The ripening kinetics of bubbles is studied by performing molecular dynamics simulations.
From the time evolution of a system, the growth rates of individual bubbles are determined.
At low temperatures, the system exhibits a $t^{1/2}$ law and the growth rate is well described by classical Lifshitz--Slyozov--Wagner (LSW) theory for the reaction-limited case. This is direct evidence that the bubble coarsening at low temperatures is reaction-limited. 
At high temperatures, although the system exhibits a $t^{1/3}$ law, which suggests that it is diffusion-limited, 
the accuracy of the growth rate is insufficient to determine whether the form is consistent with the prediction of LSW theory for the diffusion-limited case.
The gas volume fraction dependence of the coarsening behavior is also studied.
Although the behavior of the system at low temperatures has little sensitivity to the gas volume fraction up to 10\%,
that at high temperatures deviates from the prediction of LSW theory for the diffusion-limited case as the gas volume fraction increases. These results show that the mean-field-like treatment is valid for a reaction-limited system even with a finite volume fraction, while it becomes inappropriate for a diffusion-limited system since classical LSW theory for the diffusion-limited case is valid at the dilute limit. 
\end{abstract}

\maketitle

\section{Introduction}
A rapid change in a thermodynamic variable, such as temperature or pressure, makes a homogeneous system unstable
and phase separation occurs. Then nucleation, growth, and coarsening are observed~\cite{Gunton1983,GD1983, Onuki2002}.
These phenomena are very common and widely observed in any system involving a first-order transition.
Coarsening, also termed Ostwald ripening, is the process by which larger droplets of the second phase
become larger at the expense of smaller ones in the ambient phase.
There is a critical size in the system, and droplets larger than this size grow while smaller ones shrink. 
The basic theory of coarsening was constructed by Lifshitz and Slyozov~\cite{Lifshitz196135} and Wagner~\cite{Wagner1961}. The theory is now referred to as LSW theory.
This theory predicts two typical behaviors in the diffusion-limited and reaction-limited cases.
The former is the limit where the diffusion of the system is much slower than the reaction at the surface of the droplets, where the reaction involves the dissolution/redeposition of droplets, the evaporation/condensation of bubbles, and so forth.
Assuming the self-similarity of the distribution function,
the theory predicts that the asymptotic behavior of the critical radius is $t^{1/3}$, which is called the $t^{1/3}$ law.
The other limit is the opposite case where the reaction is much slower than the diffusion.
Then the critical radius behaves as $t^{1/2}$, which is called the $t^{1/2}$ law.

%%Added on the 3rd reply
Hohenberg and Halperin (HH) proposed a classification of universal models on the basis of the renormalization group~\cite{Hohenberg1977}. Systems are classified into universal models according to their properties.
If the order parameter is not conserved, the dynamics of the order parameter is expressed by the
time-dependent Ginzburg-Landau equation. This model is called Model A and predicts the $t^{1/2}$ law.
If the order parameter is conserved, \textit{i.e.}, the interfaces between two phases cannot move independently, 
the dynamics is expressed by the Cahn-Hilliard equation. This model is called Model B and predicts the $t^{1/3}$ law.
For the phase separation of binary liquid, we have to consider the hydrodynamic effect.
This is called Model H and predicts three different stages, $t^{1/3}$, $t$, and $t^{2/3}$ for the diffusive, the viscous, and the inertial regime, respectively~\cite{Bray1994}. The growth law for reaction-limited and diffusion-limited cases agree with the
Model A and Model B (or H), respectively. Note that the  gas-liquid phase transition of a one-component fluid
is classified as Model H in the review of HH and $t^{1/2}$ law will not appear in this context.

Since the publication of LSW theory, many experiments~\cite{Peng1998, Oskam2003, Zeshan2005, Viswanatha2007} as well as numerical simulations~\cite{Koch1983, Yamamoto1994, Kabrede2006, Rinald2012} have been performed.
Although the qualitative features of LSW theory, such as the self-similarity of distribution functions and the power-law behavior of the characteristic length, have been confirmed, it is widely known that the form of the distribution function deviates from that predicted by the theory~\cite{Voorhees1985,Marder1987, Oxtoby1992,Baldan2002}.
Since the original LSW theory is a mean-field theory, the many-body effect is ignored, and this approximation is justified for the dilute limit, \textit{i.e.}, the volume fraction of the second phase is negligible compared with the ambient phase.
Therefore, considerable effort has been made to extend LSW theory to a diffusion-limited system with a finite-volume fraction~\cite{Brailsford1976, Tokuyama1984, Voorhees19842001, Voorhees19842013, Enomoto1986, Brown1989}.
On the other hand, less attention has been paid to reaction-limited systems.
Viswanatha \textit{et al.} reported that the growth of ZnO lies between the diffusion- and reaction-limited cases~\cite{Viswanatha2007}. Rinaldo \textit{et al.} considered the population dynamics of Pt and discussed the reaction-limited case~\cite{Rinald2012}.
The liquid-gas phase separation has been investigated by molecular dynamics (MD) simulations, and the both $t^{1/2}$ law and
the $t^{1/3}$ law were reported depending on the conditions of simulations.
When the volume fraction of gas and liquid are comparable, then a spinodal decomposition is observed after quenching
and the $t^{1/2}$ law was observed~\cite{Yamamoto1994, Kabrede2006}.
When the volume fraction of liquid is much smaller than that of gas, on the other hand, the droplet growth is observed after quenching
and the $t^{1/3}$ law was observed~\cite{Roy2012, Jung2015}.
Koch \textit{et al.} performed isothermal and isoenergetic MD simulations~\cite{Koch1983}.
They observed the $t^{1/2}$ law for the isothermal simulation and the $t^{1/3}$ law for the isoenergetic simulation.
In our previous study~\cite{Watanabe2014}, we reported that the bubble coarsening in a decompressed liquid 
exhibits a crossover from the $t^{1/3}$ law to the $t^{1/2}$ law as the temperature increases.
These results suggest that the governing factor of the system, \textit{i.e.}, whether the system is reaction-limited or diffusion limited,
strongly depends on the conditions of the system.
However, it is difficult to determine whether the microscopic dynamics is determined by reaction or diffusion only from the macroscopic behavior such as the exponent of the average size of droplets, as pointed out by Viswanatha \textit{et al.}~\cite{Viswanatha2007}. 
Therefore, more detailed analysis is required to investigate the kinetics of bubbles.
Recently, Werz \textit{et al.}\ investigated the coarsening of particles in an Al-Cu system~\cite{Werz2014}.
Owing to improvements in the resolution of X-ray tomography, they succeeded in tracking individual particles of submillimeter size and determined the growth rate directly. They reported that the obtained growth rate significantly deviated from the prediction of LSW theory. In the case of a one-component fluid, however, it is difficult to observe the liquid-gas phase separation in experiments
since the time scale is too short and the gravity plays crucial role in expriments on Earth.
Perrot \textit{i.e.} performed experiments of phase separation in a pure fluid under microgravity environment and
observed the $t^{1/3}$ law~\cite{Perrot1994}.

Therefore, MD simulations are required to investigate the microscopic behavior of phase separation.
The simulation of bubble coarsening requires much more atoms than that of droplet coarsening.
Recent progress of computers allows us to perform MD simulations involving 38 billions of atoms~\cite{fx10full}, and therefore, MD simulations of bubble coarsening is now possible. In this study, we investigate ripening kinetics of bubbles by observing their growth rates via MD simulations involving up to 680 million particles. 

This paper is organized as follows. A brief summary of the theoretical background is given in Sec.~\ref{sec_theory}.
Section.~\ref{sec_methods} is devoted to a description of the method, particularly how to determine the growth rates of bubbles from simulation data.
The obtained results are presented in Sec.~\ref{sec_results}.
Finally, Sec.~\ref{sec_summary}~concludes the present paper with a discussion.

\section{Theory}
\label{sec_theory}

\subsection{Scaling Theory}
We start from a distribution function $f(R,t)$ that denotes the number of bubbles having radius $R$ at time $t$.
We only consider a three-dimensional system for simplicity since a two-dimensional system involves logarithmic behavior~\cite{Rogers1989, Yao1993}.
We assume that each bubble is a sphere, whose volume is given by $v = 4 \pi R^3/3$.
The time evolution of the system is governed by the following equation of continuity:
\begin{equation}
\frac{\partial f}{\partial t} = - \frac{\partial }{\partial R} \left(\dot{R} f \right), \label{eq_continuity}
\end{equation}
where $\dot{R}(R,t)$  is the kinetic equation denoting the growth rate of a bubble having radius $R$ at time $t$.
Here, we assume a mean-field-like nature, \textit{i.e.}, all bubbles in the system are subjected to an identical ambient pressure.
This assumption corresponds to the fact that the kinetic equation is single-valued.
We also assume that the kinetic equation is a continuous function, which means that there are no discontinuous jumps in the volume of bubbles due to nucleation or coalescence.
In the coarsening process, the critical radius $R_c(t)$ is defined as the radius for which a larger grows, while a smaller bubble shrinks.
Following LSW theory, we first assume the power-law behavior of the critical radius in the late stage of coarsening to be
\begin{equation}
R_c \sim t^{\alpha},
\end{equation}
with exponent $\alpha$.
We introduce the scaling variable $\tilde{R} =R/R_c$ and assume the asymptotic behaviors
\begin{eqnarray}
f(R,t) &\sim& t^{\beta} \tilde{f}(\tilde{R}), \\
\frac{\dot{R}}{R_c} &\sim & t^{\gamma} \tilde{\dot{R}}(\tilde{R}),
\end{eqnarray}
with exponents $\beta$ and $\gamma$.
The total volume of the gas in this system $V_G$ is
\begin{equation}
V_G \sim \int R^3 f \diff R = t^{\beta + 4 \alpha} \int \tilde{R}^3 \tilde{f} \diff \tilde{R}.
\end{equation}
In the late stage of coarsening, the dynamics is dominated by the surface free energy, \textit{i.e.},
the total surface area decreases while the total volume of gas is almost conserved.
The conservation of the volume of gas leads to $\beta = - 4\alpha$.
Then the power-law behavior of the total number of bubbles $n(t)$ is expressed as
\begin{equation}
n \equiv \int f \diff R = t^{-3\alpha} \int \tilde{f} \diff \tilde{R} \sim t^{-3 \alpha}.
\end{equation}
Since the scaled functions must be the solutions of Eq.~(\ref{eq_continuity}), we have $\gamma = -1$.
Therefore, there is only one exponent $\alpha$ that determines the dynamics of coarsening.

Equation~(\ref{eq_continuity}) is rewritten with the scaled functions as
\begin{equation}
\alpha \left( 4 \tilde{f}  + \frac{\diff \tilde{f}}{\diff \tilde{R}} \right)  = \frac{\diff }{\diff \tilde{R}} \left(\tilde{\dot{R}} \tilde{f} \right).\label{eq_continuity2}
\end{equation}
We rewrite Eq.~(\ref{eq_continuity2}) as
\begin{equation}
3 \tilde{f} = \frac{\diff }{\diff \tilde{R}} \left( u \tilde{f}  \right), \label{eq_r}
\end{equation}
where
\begin{equation}
u \equiv \left( \frac{\tilde{\dot{R}}}{\alpha } - \tilde{R} \right).
\end{equation}
Then Eq.~(\ref{eq_continuity2}) is integrated to obtain~\cite{Brown1989}
\begin{equation}
\tilde{f} = \frac{1}{|u|} \exp \int_0^{\tilde{R}} \frac{3}{u} \diff{\tilde{R}}. \label{eq_f}
\end{equation}
Equation~(\ref{eq_f}) implies that the distribution function $f$ can be determined from the kinetic equation.

\subsection{Kinetic Equations}

Consider a growing bubble in a system undergoing bubble coarsening.
When the bubble grows, two processes occur, a phase transition and diffusion.
The system exhibits two typical behaviors depending on which factor dominates the dynamics, the diffusion or the phase transition. In order that the bubble grows, atoms at its surface should evaporate. Then the density of gas becomes higher than that at the center of the bubble.
When the diffusion process is much slower than the evaporation rate, the evaporation is suppressed owing to the high density at the surface. Then the diffusion determines the growth rate, and this is referred to the diffusion-limited case.
On the other hand, if the evaporation rate is much slower than the diffusion process, then the gas density in the bubble is almost homogeneous and the dynamics is determined by the evaporation rate. This case is referred to the reaction-limited case.
The kinetic equation for the growth/shrinkage rates of bubbles has a different form for diffusion-limited and reaction-limited systems~\cite{Lifshitz196135,Wagner1961, Binder1977, Bray1994, Watanabe2014}. In particular, Wagner simultaneously discussed the diffusion-limited and reaction-limited cases~\cite{Wagner1961}.
In this subsection, we give a brief discussion of the kinetic equation for bubble coarsening.

In the following, we assume that the system is in hydrostatic equilibrium, \textit{i.e.},
the Young-Laplace equation
$$
\Delta P = \frac{2 \sigma}{R}
$$
 is always satisfied, where $\Delta P$ is the pressure deference between the inside and outside of a bubble,
 $\sigma$ is the surface tension, and $R$ is the radius of the bubble.
Note that this is the static limit of the Rayleigh-Plesset equation, which describes the  inertial dynamics of a bubble. This approximation is justified when the time scale of the pressure of a liquid is much faster than the dynamics involving the chemical potential, such as diffusion and evaporation/condensation processes.

We consider a current $J_R$ at the surface of a bubble.
We assume that the system is spherically symmetric and that the current is outward and oriented normal to the surface.
The steady-state solution is
\begin{equation}
J_R = \frac{D(\rho_R - \rho_0)}{R}, \label{eq_jr}
\end{equation}
where $\rho_R$ is the density of gas at the bubble surface, $D$ is the diffusion constant, and $\rho_0$ is a constant, respectively.
If the system is diffusion-limited, \textit{i.e.}, the evaporation/condensation rate is much faster than the rate of diffusion, the gas density at the surface of the bubble is equal to $\rho_R^\mathrm{eq}$, which is the equilibrium
density at the surface of the bubble having radius $R$.
Then the growth rate of the bubble is
\begin{equation}
\dot{v} = 4 \pi R^2 J_R =4 \pi R D (\rho_R^\mathrm{eq} - \rho_0) \label{eq_vdot_d}
\end{equation}
with constant $\rho_0$.
The linearized Gibbs-Thomson equation leads to 
\begin{equation}
\rho_R^\mathrm{eq} = \rho_\infty \left(1 - \frac{\lambda}{R} \right), \label{eq_gibbsthomson}
\end{equation}
where $\rho_\infty$ is the equilibrium density of gas at the flat surface and $\lambda$ is the capillary length,
which is given by
\begin{equation}
\lambda = \frac{2 \sigma V_m}{k_\mathrm{B} T},
\end{equation}
where $V_m$ is the molar volume,  $k_\mathrm{B}$ is the Boltzmann constant, and $T$ is the temperature.
Substituting Eq.~(\ref{eq_gibbsthomson}) in Eq.~(\ref{eq_vdot_d}), we have
\begin{equation}
\dot{v} \propto  \left(\frac{R}{R_c}  -1 \right).  \label{eq_dlimit}
\end{equation}
Since $\dot{R} /R_c \sim t^{-1} \tilde{\dot{R}}$, we have $\alpha = 1/3$,
and therefore, $R_c \sim t^{1/3}$, which is called the $t^{1/3}$ law.

Next, we consider the reaction-limited case.
The evaporation/condensation rate is proportional to the difference
between $\rho_R^{\mathrm{eq}}$ and $\rho_R$ when the difference is small.
Then we have
\begin{equation}
\dot{v} = 4 \pi R^2 k (\rho_R^{\mathrm{eq}} - \rho_R)
\end{equation}
with proportional constant $k$.
In the reaction-limited case, the diffusion current is virtually zero.
Therefore, Eq.~(\ref{eq_jr}) leads to $\rho_R = \rho_0$.
Considering Eq.~(\ref{eq_gibbsthomson}), we have the kinetic equation
\begin{equation}
\dot{v} \propto  R \left(\frac{R}{R_c}  -1 \right). \label{eq_rlimit}
\end{equation}
Since $\dot{R} /R_c \sim t^{-1} \tilde{\dot{R}}$, we have $\alpha = 1/2$,
and therefore, $R_c \sim t^{1/2}$, which is called the $t^{1/2}$ law.

It is worth mentioning the behavior of the kinetic equation in the limit of $R \rightarrow 0$.
Although the shrinkage rate remains finite in the diffusion-limited case (\ref{eq_dlimit}), it becomes zero in the reaction-limited case~(\ref{eq_rlimit}). Therefore, if the shrinkage rate of a small bubble approaches zero, it is strong evidence that the dynamics is the reaction-limited.

For later convenience, we introduce the scaling variable $\tilde{v}$ given by
\begin{equation}
\tilde{v} = \frac{v}{v_c},
\end{equation}
where $v_c \equiv 4 \pi R_c^3/3$ is the critical volume and its asymptotic behavior is given by
\begin{equation}
v_c = v_0 t^{3 \alpha} \label{eq_vc}
\end{equation}
with proportional constant $v_0$.
Then the kinetic equation for the reaction-limited case (\ref{eq_rlimit}) is expressed as
\begin{equation}
\dot{v} = K v_0^{1/3} t^{1/2} \tilde{v}^{1/3} \left( \tilde{v}^{1/3} -1  \right), \label{eq_reaction}
\end{equation}
with proportional constant $K$.

\begin{table}[tb]
\begin{tabular}{ccclll}
\hline
$T$ & $N$ &  $\rho_i$ & $c$ &$\rho_e$ &$\phi$ \\
\hline
       &                        &            & 1.025 &  0.712     &0.04 \\  
0.8 & 678592512 & 0.767& 1.05    &  0.663      &0.08  \\   
       &                        &           &  1.075 &  0.617     &0.1    \\ 
\hline
       &                        &            & 1.025 &  0.570     &0.04 \\  
1.0 & 542343168 & 0.613& 1.05    &  0.530      &0.06  \\   
       &                        &           &  1.075 &  0.493     &0.07    \\ 
\hline
\end{tabular}
\caption{
Simulation conditions.
The initial temperature $T$, the number of atoms $N$, the initial density $\rho_i$, 
the expansion rate $c$, the density after expansion $\rho_e$, and the volume fraction of gas $\phi$
are shown.
All systems are cubes with an initial linear size of $960$ which are expanded uniformly and adiabatically to a size of $c\times 960$.
}\label{tbl_conditions}
\end{table}

\section{Methods}
\label{sec_methods}

\begin{figure}[tb]
\includegraphics[width=8cm]{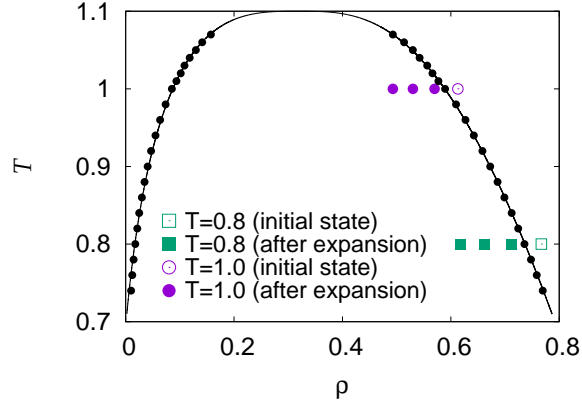}
\caption{
Phase diagram of LJ atoms with truncation.
The solid line denotes the coexisting curve of gas and liquid phases~\cite{ljuniv}.
The open symbols denote the initial states and the filled symbols denotes the states after expansion.
The square symbols correspond to simulations at lower temperature $T=0.8$
and the circle symbols correspond to those at higher temperature $T=1.0$. 
}
\label{fig_gl}
\end{figure}

\begin{figure}[tb]
\includegraphics[width=7cm]{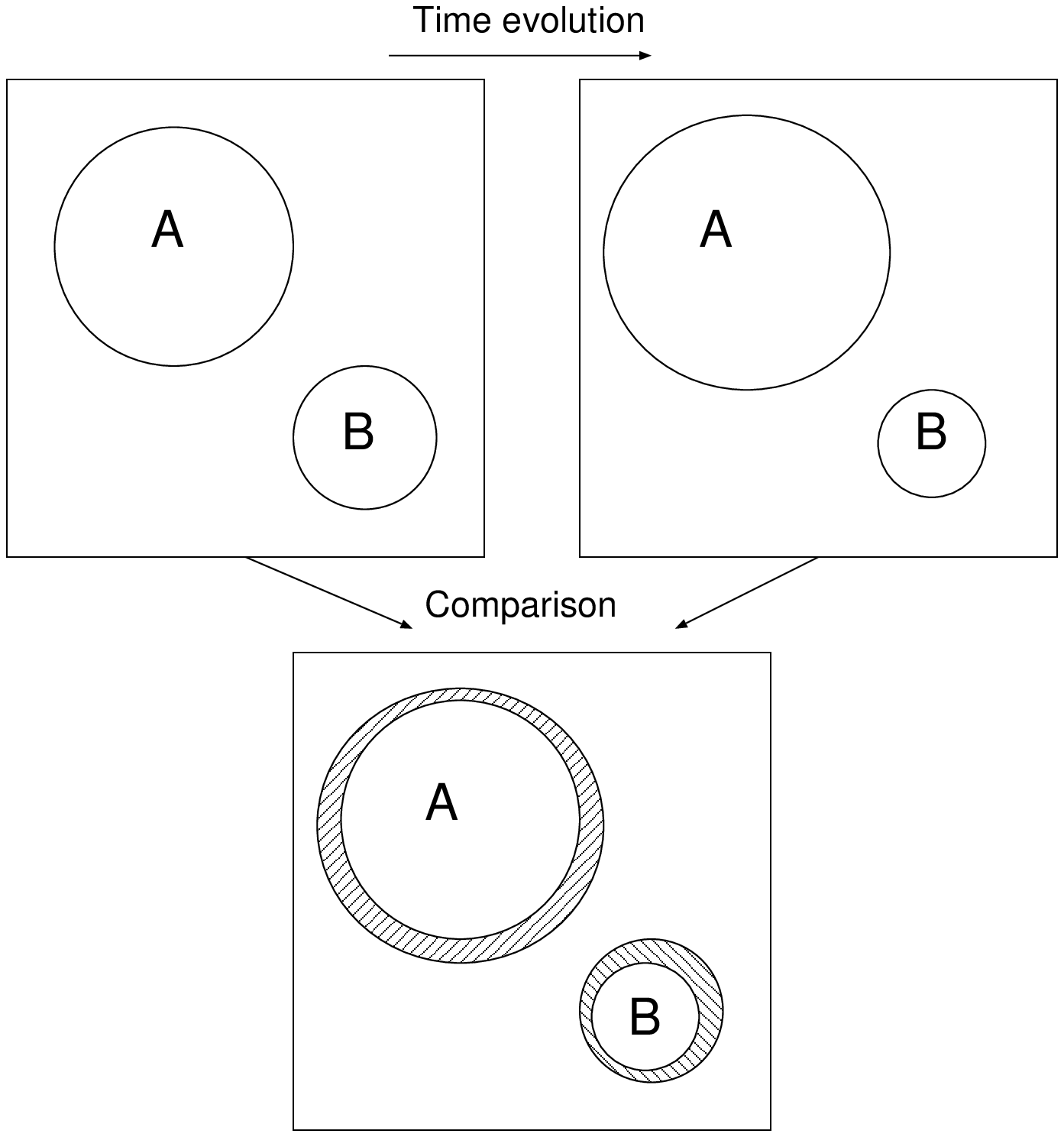}
\caption{ Schematic illustration of the calculation of the growth rate.
Showing two successive snapshots. We first identify bubbles in each snapshot and then identify which bubbles in the snapshots correspond to each other. In this example, two bubbles are found, A and B. After some time, bubble A has grown while bubble B has shrunk. We can calculate the growth and shrinkage rates of the bubbles by a finite difference approximation.
}
\label{fig_bubble}
\end{figure}

We performed MD simulations with the truncated Lennard-Jones (LJ) potential. Refer to the previous study for details of the simulation~\cite{Watanabe2014}. In the following, we measure the physical quantities in the units of LJ, \textit{i.e.}, the Boltzmann constant $k_\mathrm{B}$ is set to unity, the length is measured in terms of the diameter of  atoms, and so forth.
The system is a cube with linear size 960.
The periodic boundary condition is taken in all directions.
The time step is fixed to $0.005$ throughout the simulations.
The system is first thermalized to the pure-liquid phase using the Nos\'e--Hoover thermostat~\cite{Hoover1985}.
The time evolution of the isothermal system is performed using the reversible reference system propagation algorithm (r-RESPA)~\cite{Tuckerman1992}.
We consider two initial temperatures, $T=0.8$ and $1.0$.
After thermalization of $10^4$ steps, the thermostat is turned off and uniform and adiabatic expansion is performed.
The linear size of the system is then changed from $L$ to $cL$ with expansion rate $c$.
The linear size of the system before expansion is set to $960$.
We perform the expansion with three values of the expansion rate, $c = 1.025, 1.05,$ and $1.075$.
The parameters used in the simulations are summarized in Table~\ref{tbl_conditions}.
The phase diagram and the state points simulated in the present study are shown in Fig.~\ref{fig_gl}.
Note that, the temperature decreases instantly after expansion and gradually increases during coarsening, which is not reflected
in the figure.

The initial density and the smallest value of the expansion rate are chosen so that
the system after expansion becomes unstable and immediately exhibits spinodal decomposition~\cite{Watanabe2010}.

Snapshots of the system are stored every $1000$ steps. 
After the simulations, we identify the bubbles in each snapshot by the subcell-dividing method~\cite{Warren2001,Watanabe2010, fx10full, Watanabe2014}, in which the system is divided into subcells and the local density is computed in each subcell.
A subcell is identified as being in the gas phase when its local density is lower than some threshold. Neighboring subcells in the gas phase are identified as being in the same bubble. Since the densities of a gas and liquid differ substantially, the identification process is robust against the value of the threshold.
We choose the linear length of subcells to be about $3$, which determines the resolution of the bubbles volume.
Simulations are performed with a parallelized MD program~\cite{mdnote, mdacp}.
After expansion, the volume fraction of the gas phase to the liquid phase $\phi$ becomes almost constant during the coarsening.
The values of $\phi$ are also shown in Table~\ref{tbl_conditions}.

We calculate the growth or shrinkage rate of bubbles as follows.
Consider two successive snapshots at times $t - \tau$ and $t +\tau$.
First, we identify the bubbles in each snapshot. If two bubbles have a spatial overlap across two snapshots, then they are considered to be the same bubble (see Fig.~\ref{fig_bubble}). Suppose bubble $i$ has volume $v_i$ at time $t$.
To obtain the growth rate of the bubble, we adopt the central difference
\begin{equation}
\dot{v_i}(t) \sim \frac{v_i(t+\tau) -v_i(t-\tau)}{2 \tau},
\end{equation}
with a fixed time interval $\tau$. 
In this simulation, we use $\tau = 5$.
A pair comprising the volume and the growth rate $(v(t),\dot{v}(t))$ can be determined from each bubble in the snapshots.
We calculate the growth rates of all the bubbles in a snapshot at time $t$. The set of growth rates is simply the kinetic equation.

\section{Results} \label{sec_results}
\begin{figure}[tb]
\includegraphics[width=8cm]{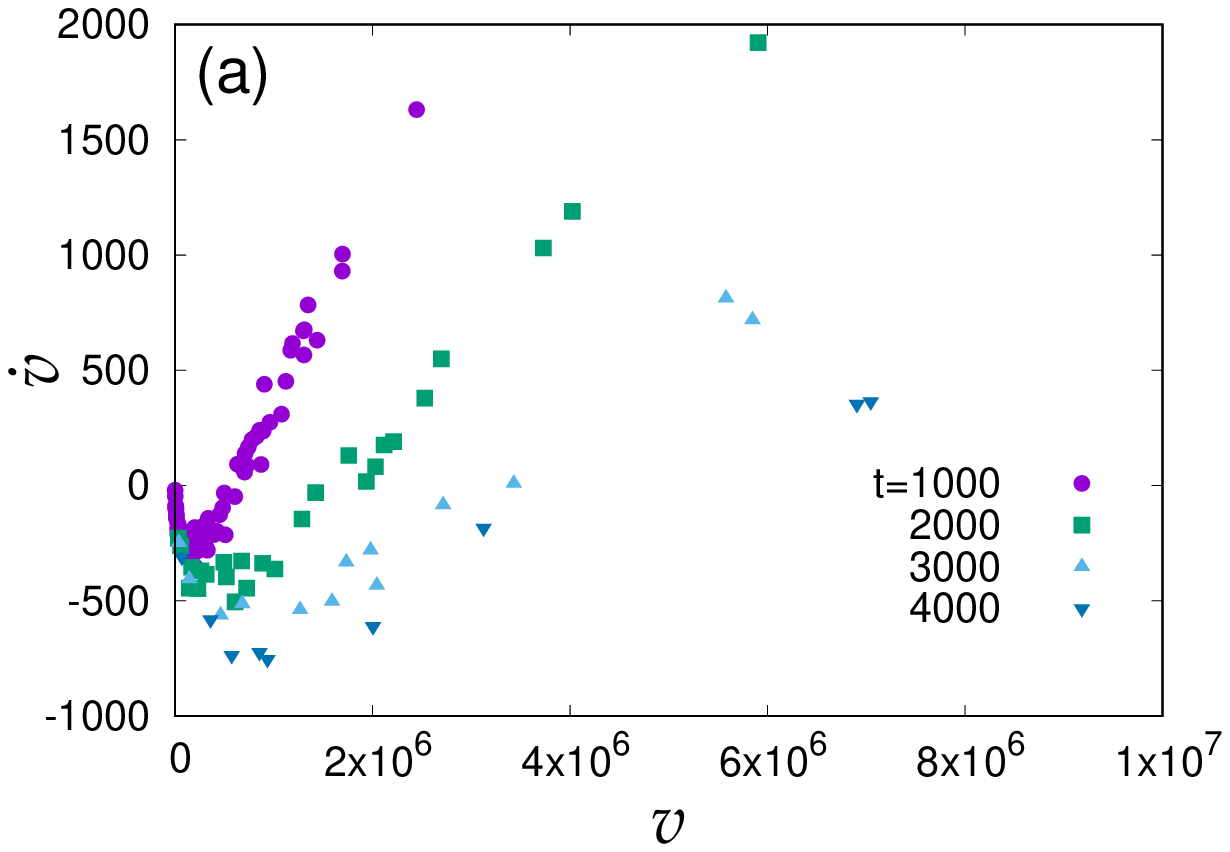}
\includegraphics[width=8cm]{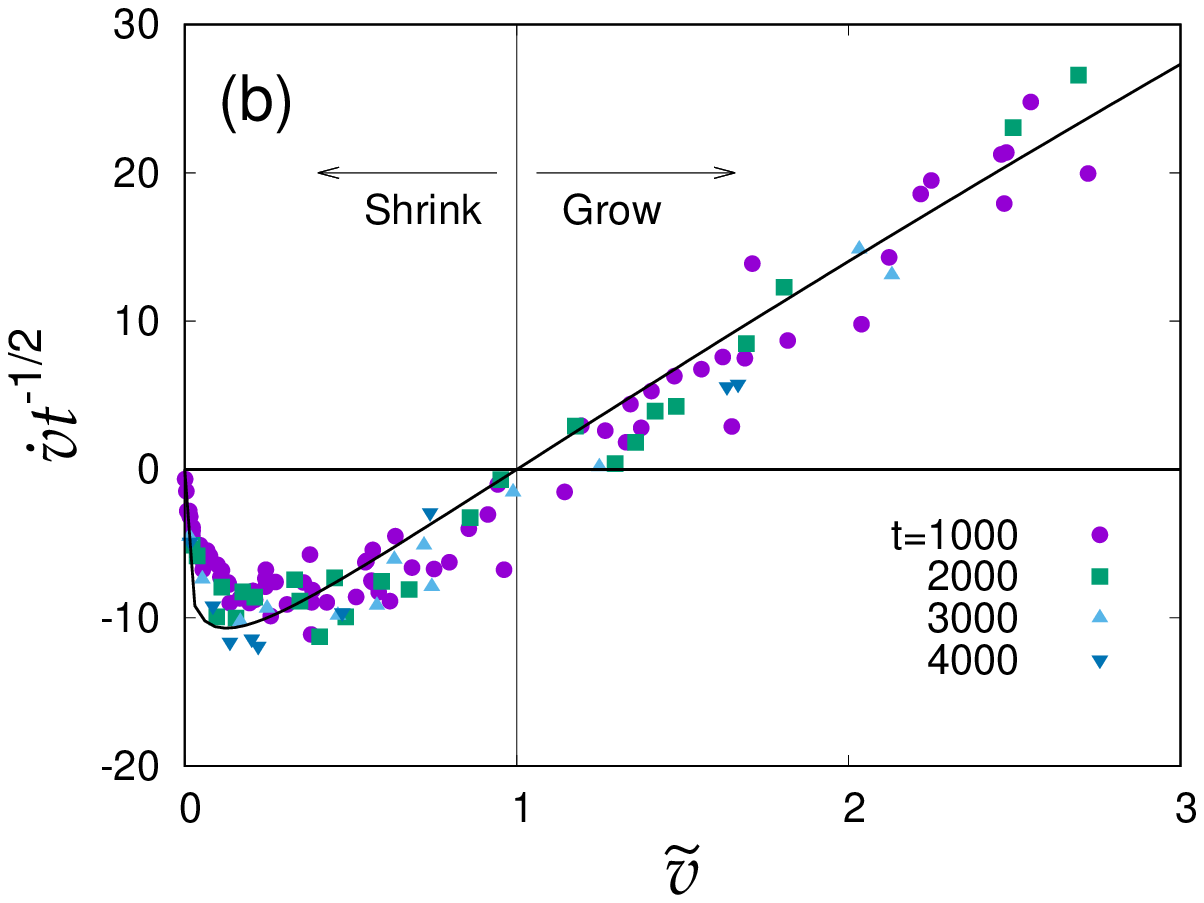}
\caption{
(Color online) (a) Growth rates of bubbles at $t=1000, 2000, 3000,$ and $4000$. The temperature is $0.8$ and the gas fraction is $\phi =0.04$.
 (b) Scaling plot of the growth rates. We assume $\alpha = 1/2$. The solid line denotes the theoretically predicted form given by Eq.~(\ref{eq_reaction}) with $K= 16.8(4)$ and $v_0 = 16.7(3)$.
} \label{fig_vdot}
\end{figure}

\begin{figure}[tb]
\includegraphics[width=8cm]{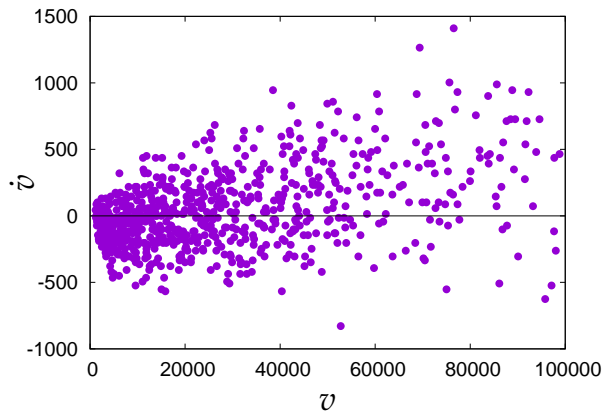}
\caption{
Growth rates of bubbles at $t=1000$ for the system with $T= 1.0$ and $\phi =0.04$.
It is difficult to determine whether the growth rate is of the form given by Eq.~(\ref{eq_dlimit}).
} \label{fig_higher}
\end{figure}

\subsection{Scaling Behavior of the Kinetic Term}

The obtained growth rates for the case of  $T=0.8$ and $\phi=0.04$ are shown in Fig.~\ref{fig_vdot}~(a).
The growth rates are found to be almost single-value functions of $\tilde{R}$.
This means that the bubbles are subjected to an identical pressure at each time, and therefore, a mean-field-like treatment is justified at this temperature.
A scaling plot of the growth rates is shown in Fig.~\ref{fig_vdot}~(b).
The growth rates at different times are well scaled with respect to the scaling variable $\tilde{v}$.
One can also confirm that there is a critical volume of $\tilde{v}=1$, and the growth rates of bubbles smaller than the critical volume are negative, and therefore, the smaller bubbles shrink and the larger bubbles grow.
The solid line is the theoretical prediction given by Eq.~(\ref{eq_reaction}).
There are two fitting parameters, $K$ and $v_0$, which are determined to be $16.8(4)$ and $16.7(3)$, respectively.
The function form of the growth rate is well described by the prediction for the reaction-limited case~(\ref{eq_rlimit}) and is clearly different from that in the diffusion-limited case~(\ref{eq_dlimit}). 

In our previous work, we observed a crossover from the $t^{1/2}$ law to the $t^{1/3}$ law as the temperature increased~\cite{Watanabe2014}. Considering the value of the exponent, the system with $T=1.0$ is expected to diffusion-limited.  However, we found that it is difficult to determine whether the growth rate has the form described in Eq.~(\ref{eq_dlimit}), as shown in Fig.~\ref{fig_higher}.

\subsection{Critical Volume}

\begin{figure}[tb]
\includegraphics[width=8cm]{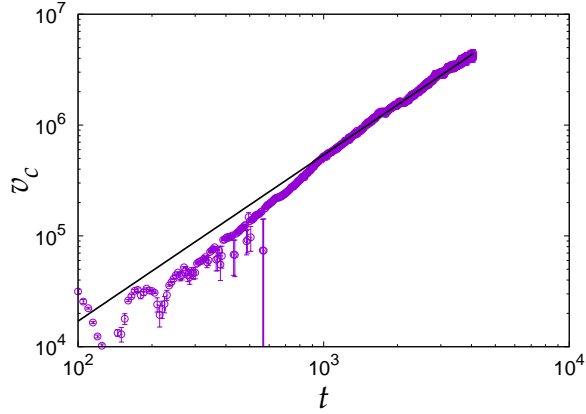}
\caption{
(Color online) Time evolution of the critical volume at $T=0.8$ and $\phi=0.04$.
Each point is obtained by fitting Eq.~(\ref{eq_reaction}) to the data of the kinetic term.
The fitting errors are smaller than the size of the symbols in most points.
The solid line is $v_0 t^{1.5}$ and $v_0$ is estimated to be $17.02(3)$. 
Decimal logarithms are taken for both axes.
} \label{fig_vc}
\end{figure}

Since the function form of the growth rate is well described by the theoretical prediction, we can estimate the critical volume at each time by fitting Eq.~(\ref{eq_reaction}) to the obtained data.
For the fitting, we fix the value of the coefficient $K$ to $16.8$, which is obtained from the scaling behavior of the growth rates.
Therefore, the only fitting parameter is the critical volume $v_c(t)$ at each time $t$.
The time evolution of the critical volume at $T=0.8$ is shown in Fig.~\ref{fig_vc}.
In the scaling regime ($t>10^3$), the critical volume is well described by the power-law behavior $t^{3\alpha}$ with $\alpha =1/2$. Assuming Eq.~(\ref{eq_vc}), the coefficient of the critical volume is estimated to be $v_0 = 17.02(3)$, which is consistent with the value of $v_0 = 16.7(3)$ obtained from the scaling behavior.

\subsection{Distribution Function}

\begin{figure}[tb]
\includegraphics[width=8cm]{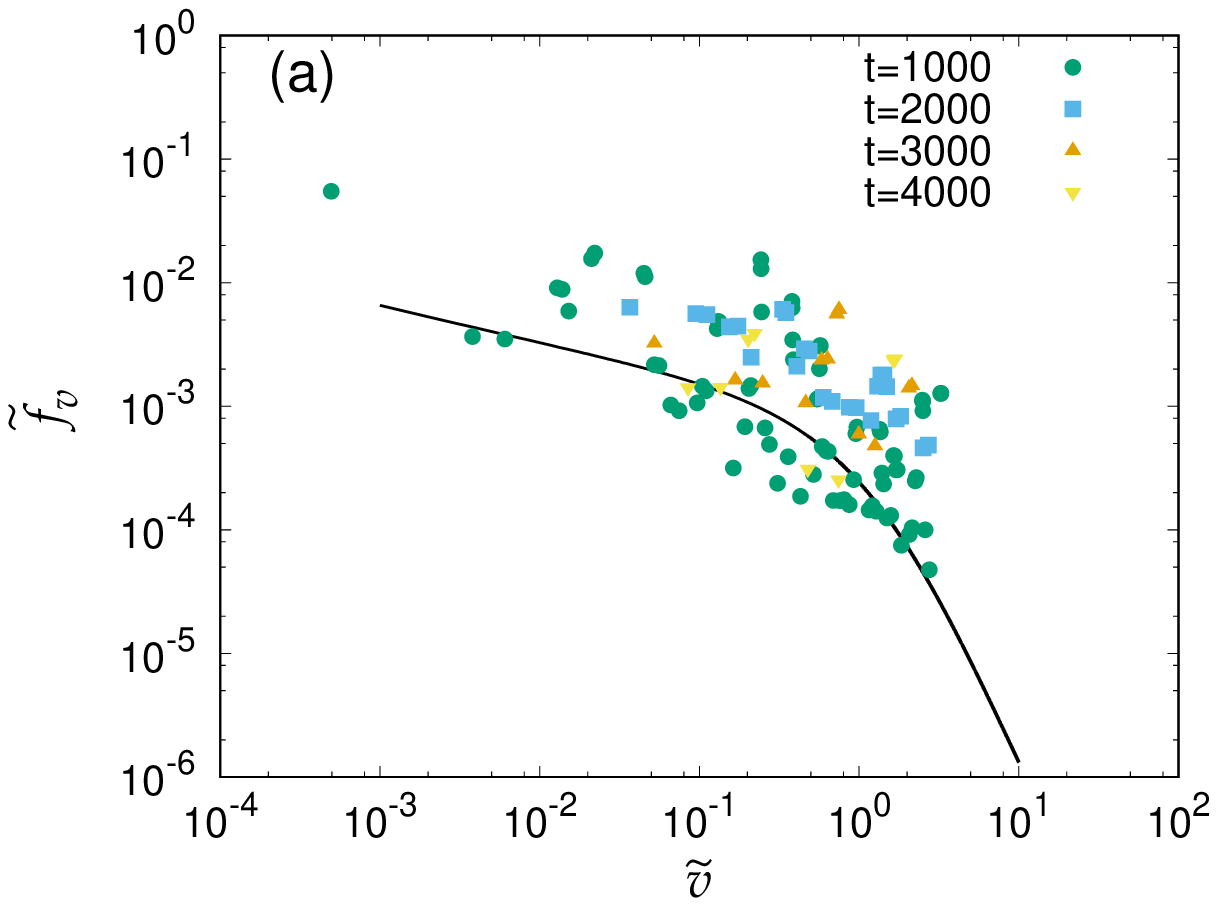}
\includegraphics[width=8cm]{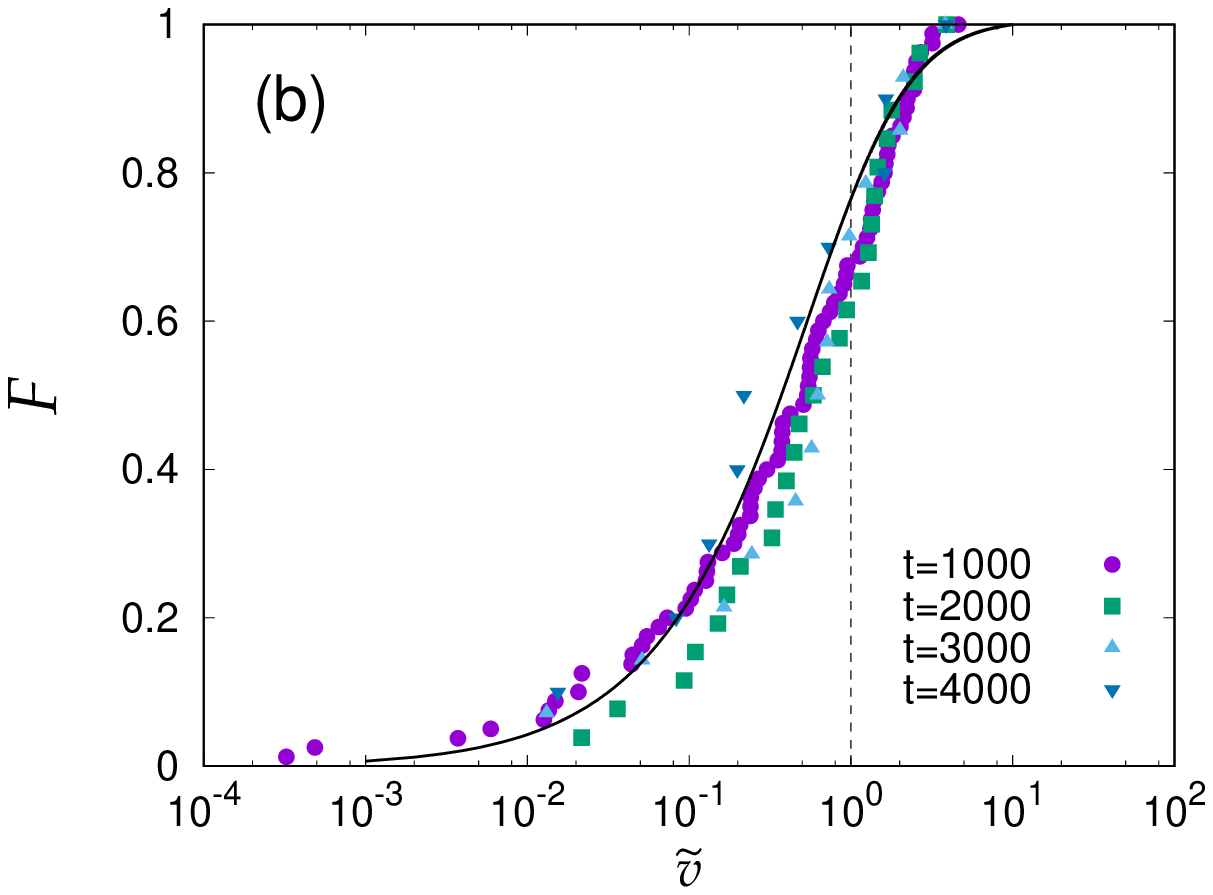}
\caption{
(Color online) Scaling plots of (a) distribution functions $\tilde{f}_v$ and (b) cumulative distribution functions $F(\tilde{v})$.
The data at $t=1000, 2000, 3000$, and $4000$ are shown.
The solid line denotes the reconstructed distribution function and the cumulative distribution function
using Eq.~(\ref{eq_f}) with $K=16.8$ and $v_0 = 17.02$.
The decimal logarithm is taken for both axes in (a) and for horizontal axis in (b).
} \label{fig_cdf}
\end{figure}

The kinetic equation~(\ref{eq_reaction}) contains only two coefficients, $K$ and $v_0$.
Since we have determined their values, we can reconstruct the distribution function using Eq.~(\ref{eq_f}).
The scaling behavior of the distribution functions $\tilde{f}_v$ and the cumulative distribution functions (CDFs) $F(\tilde{v})$ at $T=0.8$ are shown in Fig.~\ref{fig_cdf}, where $\tilde{f}_v(\tilde{v}) \equiv 4 \pi \tilde{R}^2$ is the distribution function of the volume and $F(\tilde{v})$ is its CDF, which is given by 
\begin{equation}
F(\tilde{v}) = \int_0^{\tilde{v}} \tilde{f}_v \diff \tilde{v}.
\end{equation}
Although the CDFs are well scaled using the scaling variable $\tilde{v}$, the statistical precision is insufficient to discuss whether the form of the distribution function deviates from the classical theory.

\subsection{Gas Volume Fraction Dependence}

\begin{figure}[tb]
\includegraphics[width=8cm]{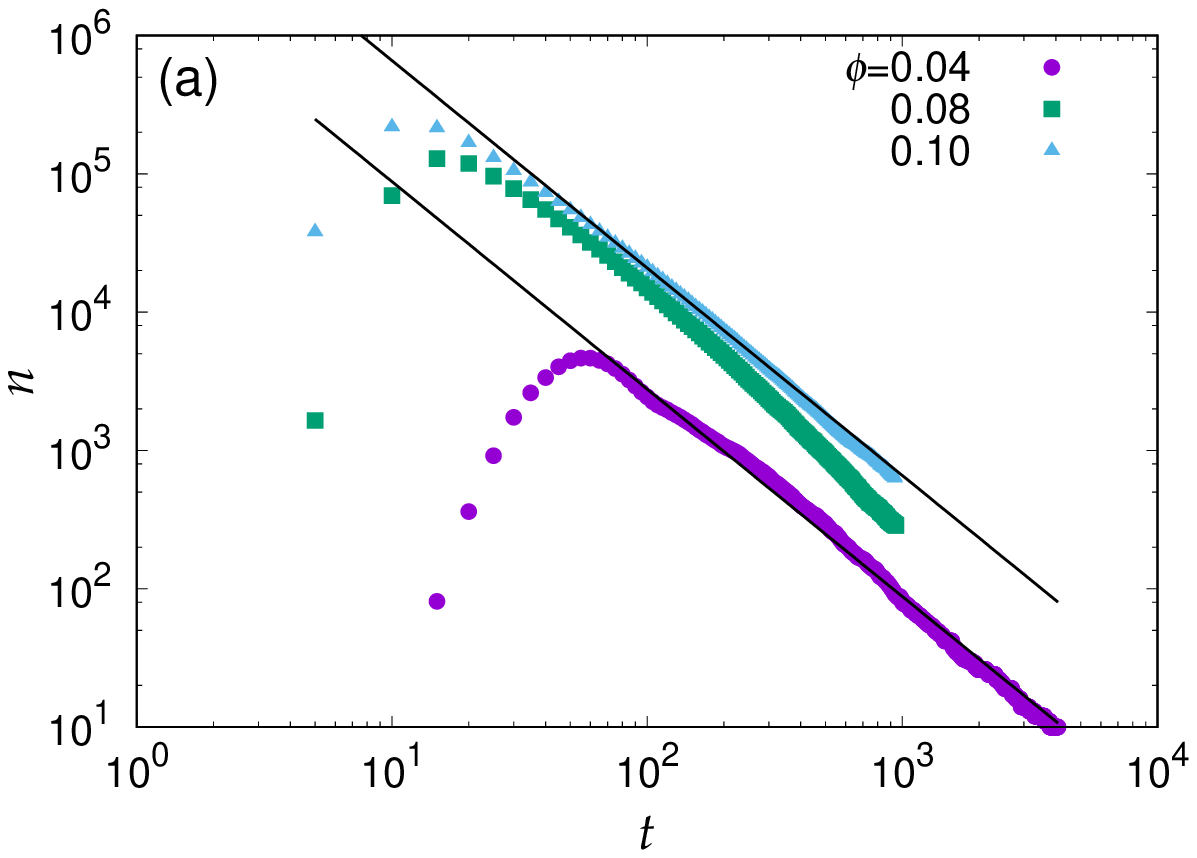}
\includegraphics[width=8cm]{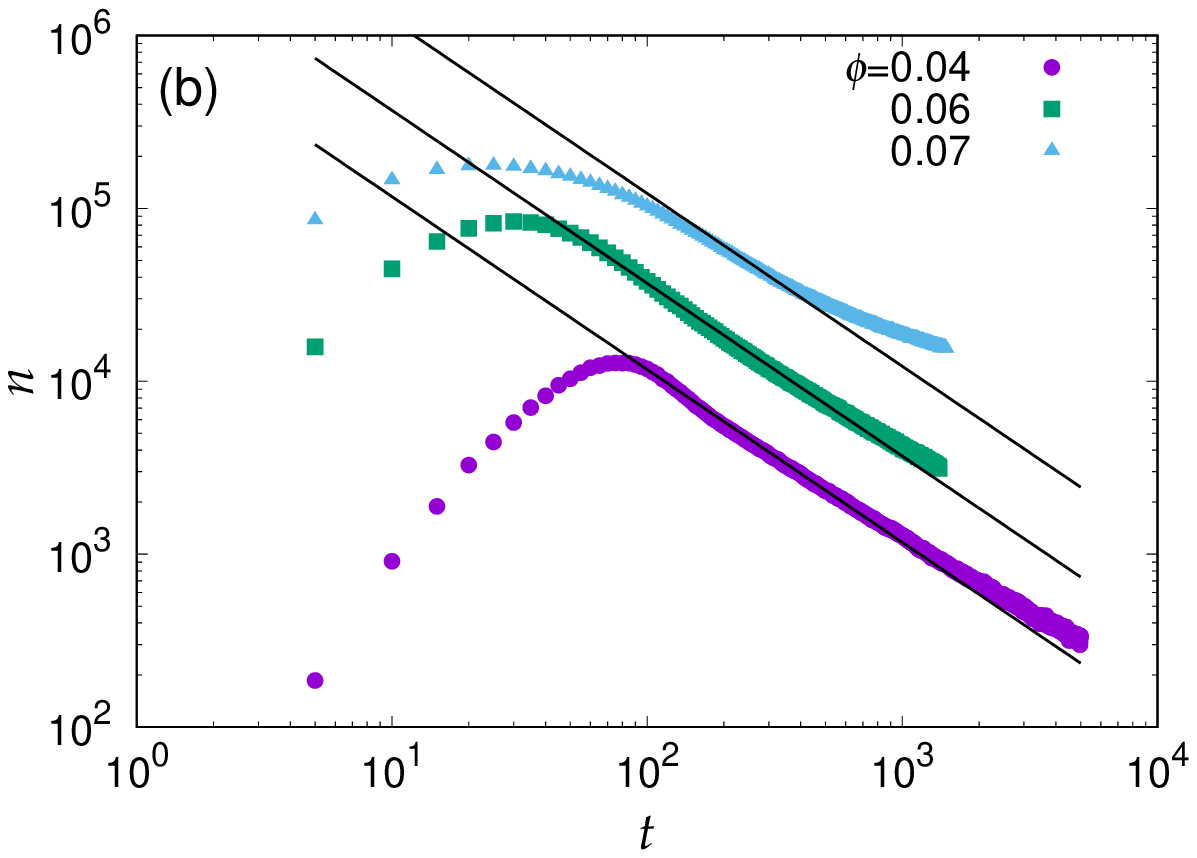}
\caption{
(Color online) Time evolutions of the total number of bubbles.
The decimal logarithm is taken for both axes.
(a) Data at a low temperature of $T=0.8$.
The solid lines $t^{-1.5}$ are a guide to the eyes.
(b) Data at a high temperature of $T=1.0$.
The solid lines $t^{-1}$ are a guide to the eyes.
As the volume fraction of the gas increases, the power-law behavior departs
from the diffusion-limited case with $n \sim t^{-1}$.
} \label{fig_bnumber}
\end{figure}

In this subsection, we consider the gas volume fraction $\phi$ dependence of the coarsening behavior.
By changing the rate of expansion, we can control the gas volume fraction during coarsening.
Since the initial temperature and density are fixed, larger values of $\phi$ correspond to 
deeper quenching.
The temperature and gas volume fraction dependences of the time evolution of the total number of bubbles
are shown in Fig.~\ref{fig_bnumber}.
At a low temperature, the power-law behavior is found to be insensitive to the gas volume fraction.
This is consistent with the fact that the system is reaction-limited at a low temperature.
If the system is reaction-limited, the dynamics is governed by the evaporation/condensation at the surface of the bubbles,
and therefore, the dynamics is less sensitive to the volume fraction of the second phase.

Meanwhile, the behavior at a high temperature is found to be sensitive to the gas volume fraction.
Although the power-law behavior of the total number of bubbles is well described 
by the diffusion-limited case with $n \sim t^{-1}$, it departs from it as the volume fraction of the gas increases.
This is consistent with the fact that the system is diffusion-limited at a high temperature.
Since the classical LSW theory for diffusion-limited case is justified only for the dilute limit,
the behavior of the system deviates from the theoretical prediction as the volume fraction of the gas phase increases.

\section{Summary and Discussion}
\label{sec_summary}
We have performed molecular simulations of bubble coarsening.
We observed both $t^{1/2}$ and $t^{1/3}$ laws which correspond to
the reaction-limited and the diffusion-limited cases of LSW theory, respectively.
From the time evolution of the bubble configurations, we directly determined the growth rates of the bubbles.
At low temperatures, the function form of the growth rate is consistent with that predicted by LSW theory for the reaction-limited case.
The scaling exponent was found to have little sensitivity to the gas volume fraction up to 10\%.
These results are consistent with LSW theory for the reaction-limited case.
When the system is reaction-limited, the diffusion processes are negligible, the pressure of the ambient liquid is almost homogeneous, and the dynamics of bubble coarsening is determined only by that at the surfaces of the bubbles.
Therefore, a mean-field-like treatment is justified even for a finite volume fraction of the gas phase.
Although the scaling behaviors of CDFs and growth rates are clearly confirmed, the accuracy of the results is insufficient to discuss the shape of the form. This should be examined in further studies.

At high temperatures, the total number of bubbles behaves as $n\sim t^{-1}$, which suggests that the system is diffusion-limited, and the power-law behavior departs from this form as the volume fraction of gas increases.
These results are consistent with the fact that classical LSW theory for the diffusion-limited case is valid for the dilute-limited case. 

When we performed MD simulations involving about $10^{8}$ atoms, there are up to $10^5$ bubbles in the simulation box. However, the accuracy of the distribution function was insufficient for further analysis.
Although the growth rates at low temperatures were determined with acceptable accuracy, the accuracy at high temperatures was insufficient to determine whether the form was that predicted in the theory.
Larger simulations are required to improve the accuracy.

\section*{Acknowledgements}
The computations were carried out on the K computer provided by the RIKEN Advanced Institute for Computational Science through the HPCI System Research project (Project ID:hp130047 and hp160267) and on the facilities of the Supercomputer Center, Institute for Solid State Physics, University of Tokyo.
We would like to thank H. Hayakawa and N. Kawashima for helpful comments.
This work was supported by JSPS KAKENHI Grant Numbers 23740287 and 15K05201.

\bibliography{kinetic}

\end{document}